\documentclass[11pt]{article}
\usepackage{graphicx}
\usepackage{epstopdf} %needs to be loaded after graphicx
\usepackage{helvet}
\usepackage{setspace}
\usepackage{rotating}
\usepackage{booktabs}
\usepackage{lscape}
\usepackage[superscript]{cite}
\usepackage{caption}
\usepackage{longtable}
\usepackage{color}
\begin{document}
\thispagestyle{empty}
\fontfamily{phv}\selectfont{%Helvetica
%{\huge \noindent Non-aqueous Lithium-air battery electrolyte solvents: \\
%Charting the known chemical space}
{\huge \noindent Charting the known chemical space for non-aqueous Lithium-air battery electrolyte solvents}
\\
\\
{\scriptsize Tamara Husch, Martin Korth$^{*}$\\
%$^{1}$
Institute for Theoretical Chemistry, Ulm University, Albert-Einstein-Allee 11, 89069 Ulm (Germany)\\
$^{*}$ Corresponding author email: martin.korth@uni-ulm.de}\\ \\

%\noindent {\bf \fontfamily{phv}\selectfont{Keywords:}} electrolytes, molecular organic materials, screening, computational chemistry
%, solid-electrolyte-interphase, graphite exfoliation\\ \\

\noindent {\bf \fontfamily{phv}\selectfont{Abstract}}\\
\noindent The Li-air battery is a very promising candidate for powering future mobility,
but finding a suitable electrolyte solvent for this technology turned out to be a major problem.
We present a systematic computational investigation of the known chemical space for possible Li-air electrolyte solvents.
It is shown that the problem of finding better Li-air electrolyte solvents is not only -- as previously suggested --
about maximizing $Li^+$ and $O_2^-$ solubilities, but about finding the optimal balance of these solubilities with the viscosity of the solvent.
As our results also show that trial-and-error experiments on known chemicals are unlikely to succeed,
full chemical sub-spaces for the most promising compound classes are investigated,
and suggestions are made for further experiments.
The proposed screening approach is transferable and robust and can readily be applied to optimize electrolytes for other electrochemical devices.
It goes beyond the current state-of-the-art both in width (considering the number of compounds screened and the way they are selected),
as well as depth (considering the number and complexity of properties included).
\\

\noindent {\bf \fontfamily{phv}\selectfont{Introduction}}\\
\noindent Li-ion batteries have enabled the success of mobile electronic devices, but are not yet suited for competitive application in electric vehicles.
Recent years have accordingly seen a tremendous amount of work devoted to go beyond standard Li-ion intercalation technology.
%HERE
Especially the Li-air battery holds great promise, as it has the highest theoretical energy density of all Lithium-based alternatives,
and many researchers all over the world now investigate its chemistry.\cite{Bhatt2014, Luntz2014, Lu2014} 
In the Li-air battery, $O_2$ enters the cathode on discharge, where it is reduced and reacts with $Li^+$ ions to form $Li_2O_2$.
Upon charging, the $Li_2O_2$ is oxidized to evolve oxygen.
Though the Li-air technology still faces many different challenges, the selection of a suitable electrolyte
for the reactive environment of the oxygen cathode has been identified as one of the key obstacles.\cite{Balaish2014} 
Very recently, Luntz and coworkers as well as Bruce and coworkers identified the solubilites of $Li^+$ and $O_2^-$ as crucial parameters
on the basis of new detailed insight into the mechanisms causing Li-air batteries to die prematurely. \cite{Johnson2014,Aetukuri2015}
Dimethylsulfoxide (DMSO) and Methyl-Imidazol (MeIm) were shown to offer substantial improvement over conventional electrolytes,
but a search for better choices was encouraged in both studies, as for instance DMSO is not stable as a long term electrolyte.\cite{Kwabi2014}

The identification of new Li-air electrolyte solvents was targeted in the past (see below for details and references).
These studies investigated compounds or compound families, that were hand-picked based on previously reported desirable properties or chemical intuition.
As the chemical space of possible polar-aprotic, organic liquids is tremendously large, 
a rational decision-making model which compounds to investigate experimentally is highly desirable.
The mere vastness of the space under consideration does make it seem very likely,
that improving upon the current solvents is possible.
At least today it seems impossible to test compounds experimentally in a magnitude that allows systematic investigations in this sense,
but we will show in the following that computational high-throughput screening now offers a way to probe the full known chemical space
and make systematic investigations of full sub-spaces possible.
Screening should be seen as complementary to detailed experimental and computational investigations,
as it first requires detailed insight into the relevant processes to identify suitable screening parameters,
but offers then a way to transfer insight into innovation by reducing effort on trial-and-error procedures through rational pre-selection.

Large-scale computational screening in battery research was first applied to identify new inorganic materials
for cathodes by Ceder and coworkers within the Materials Project.\cite{Jain2013}
Other fields of renewable energy research have seen similar investigations.
A prominent example is the Harvard Clean Energy Project that strives to identify organic molecules for photovoltaics.\cite{Hachmann2014} 
The scope of electronic structure theory based screening projects in renewable energy research reaches from a few thousands to a few million
(Materials Project: 60K,\cite{Jain2013} Harvard Clean Energy Project: 2.3 Mio \cite{Hachmann2014}).
The utilization of screening techniques to optimize battery electrolytes is still in its early stages.
Several exploratory studies with a strong focus on redox stabilities were published in the past.\cite{Korth2015}
Only this year the groups of Korth and shortly afterwards Curtiss published larger scale studies based on more properties than just redox stabilities.\cite{Husch2015, Cheng2015}

Another noticeable feature of all published studies is the choice of the structural pool. The structural pool consists
 of known electrolyte molecules or the candidates are derived from a given motif, that is identified from experimental insight or chemical intuition.
The chemical space under investigation thus suffers from a 'selection bias',\cite{MB}
and the question of how to navigate chemical space needs to be addressed to alleviate the effects of this bias.
First steps in this direction were made by us when evaluating computational methods at different theoretical levels
for the identification of new battery electrolyte solvents.\cite{Korth2014,Husch2015}
Here we chart the known chemical space represented by the largest publicly available database,
to identify promising candidates and relevant structural motifs for new Li-air battery electrolyte solvents.
As a second step we systematically investigate full sub-spaces for the most promising compound classes.

Our screening methodology itself goes beyond the current state-of-the-art by 
including computational estimates for all properties reported as relevant so far.
By evaluating our data with respect to multiple properties at the same time,
many false predictions are avoided, which is of utmost importance
when making suggestions for subsequent experimental work.
A reasonably large body of knowledge on Li-air electrolyte solvents is available from both experimental\cite{Balaish2014}
and theoretical\cite{Bryantsev2013b, Bryantsev2011, Bryantsev2012, Bryantsev2011a, Bryantsev2013a, Khetan2014, Khetan2014a} investigations.
This allows us to validate our screening results for the known chemical space in the first part,
thereby giving support to our suggestions for new compounds in the second part.\\

\noindent {\bf \fontfamily{phv}\selectfont{Screening Protocol}}

\noindent Relevant screening parameters have been collected by analyzing the literature on Li-air battery electrolyte solvents.
Some requirements for a suitable electrolyte are inherited from Li-ion technology:
High electrochemical stabilities, suitable melting and boiling points, high flash points,
low viscosities/high ion conductivities, and high ion solubilities (as well as low toxicity and cost).\cite{Xu2014}
Estimates for these properties can be computed with quantum chemical methods and the COSMOtherm model.\cite{Klamt2011}
Recently, the performance of COSMOtherm for the relevant properties was evaluated
on a set of standard electrolyte solvents and typical errors of about 5-10\% were found.\cite{Husch2015}
More importantly for our case, Pearson R values for the correlation of theoretical predictions
with experimental measurements are very high, thus indicating that COSMOtherm is very well suited
for ranking compounds with respect to these properties.
Additional criteria have to be met in the case of Li-air batteries,
like high oxygen solubilities and diffusivities.
Especially Khetan {\it et al.},\cite{Khetan2014, Khetan2014a} and Bryantsev {\it et al.}
\cite{Bryantsev2013b, Bryantsev2011, Bryantsev2012, Bryantsev2011a, Bryantsev2013a}
have contributed greatly to identifying suitable descriptors for Li-air battery electrolyte solvents.
Their work emphasizes the importance of the chemical stability of the solvent in the rough oxygen cathode environment,
where it is subjected to strong bases and nucleophiles like the superoxide anion $O_2^{\bullet -}$.
Bryantsev {\it et al.} showed that the pK$_a$ of the solvent is a reasonable estimator for the stability towards superoxide,
\cite{Bryantsev2011a, Bryantsev2013b, Bryantsev2012, Bryantsev2013a} which additionally mediates autooxidation\cite{Bryantsev2012},
so that solvents with high pK$_a$s should also be more unlikely to undergo autoxidation.

Very recent results by Johnson {\it et al.} highlight the importance of good $Li^+$ solubilities,
as they are related to changes in the morphology of the $Li_2O_2$ discharge product:\cite{Johnson2014}
Solvents with poor $Li^+$ solubility lead to $Li_2O_2$ film growth that is associated with low capacity, decaying rates and early cell death.
In contrast, solvents with good $Li^+$ solubility lead to particle growth, a higher capacity and sustained discharge.
Shortly afterwards Luntz and coworkers reached the same conclusion, additionally emphasizing the importance of the $O_2^-$ solubility.\cite{Aetukuri2015}
According to these studies, the problem is thus (to first approximation) two-dimensional:
Solvents with high solubilities for both $Li^+$ and $O_2^-$should allow for high-capacity Li-air batteries.
Both studies are based on quantifying ion solubility with Gutman donor (for cations) and/or acceptor (for anions) numbers (AN and DN).
The COSMOtherm model allows to compute solubilities based on input from quantum chemical calculations,
thus providing an alternative that is suitable for large scale computational screenings.
%\textcolor{red}{ %ALERT
As high solubilities are indicated by zero by the COSMOtherm model, we turn to the chemical potential as defined in the COSMO-RS theory \cite{Klamt2011} for 
quantification.
%
%\begin{flushleft}
%
%\end{flushleft}
%}
%\\
We found the correlation between the donor number and the $Li^+$ chemical potential to be high (R=0.85),
with substantial deviations only for very high donor numbers,
which in turn do not correlate well with experimental data ({\it c.f.} Supplementary Information Section 1).

Structures were obtained from the PubChem Compound database.\cite{Bolton2008}
It comprised 67 million compounds at the time of retrieval. 
The PubChem database started in 2004 as United States Government initiative and is maintained by the National Center for Biotechnology Information.
Initially it was designed to collect information on (biological) activities of small molecules,
but was since extended and many journals today automatically contribute to the extension.
The PubChem Compound database is the closest image of the known chemical space that is publicly available.
The only larger database in this field is the fee-based CAS registry (currently 91 million ), which is commonly seen as a direct competitor.

For the first stage of the screening process we propose a hierarchical down-selection strategy as illustrated in figure \ref{fig:screeningstrategy}
(Arguments for the validity of this strategy are given in Supplementary Information section 2.)
The first steps at this stage are based on global criteria for organic, molecular electrolyte materials, i.e.
compounds are discarded that are unlikely for application in any Lithium battery technology.
In the next steps, criteria specific for Li-air battery electrolytes are applied.
 
\begin{figure}[h]
 \begin{center}
  \includegraphics[width=0.8\textwidth]{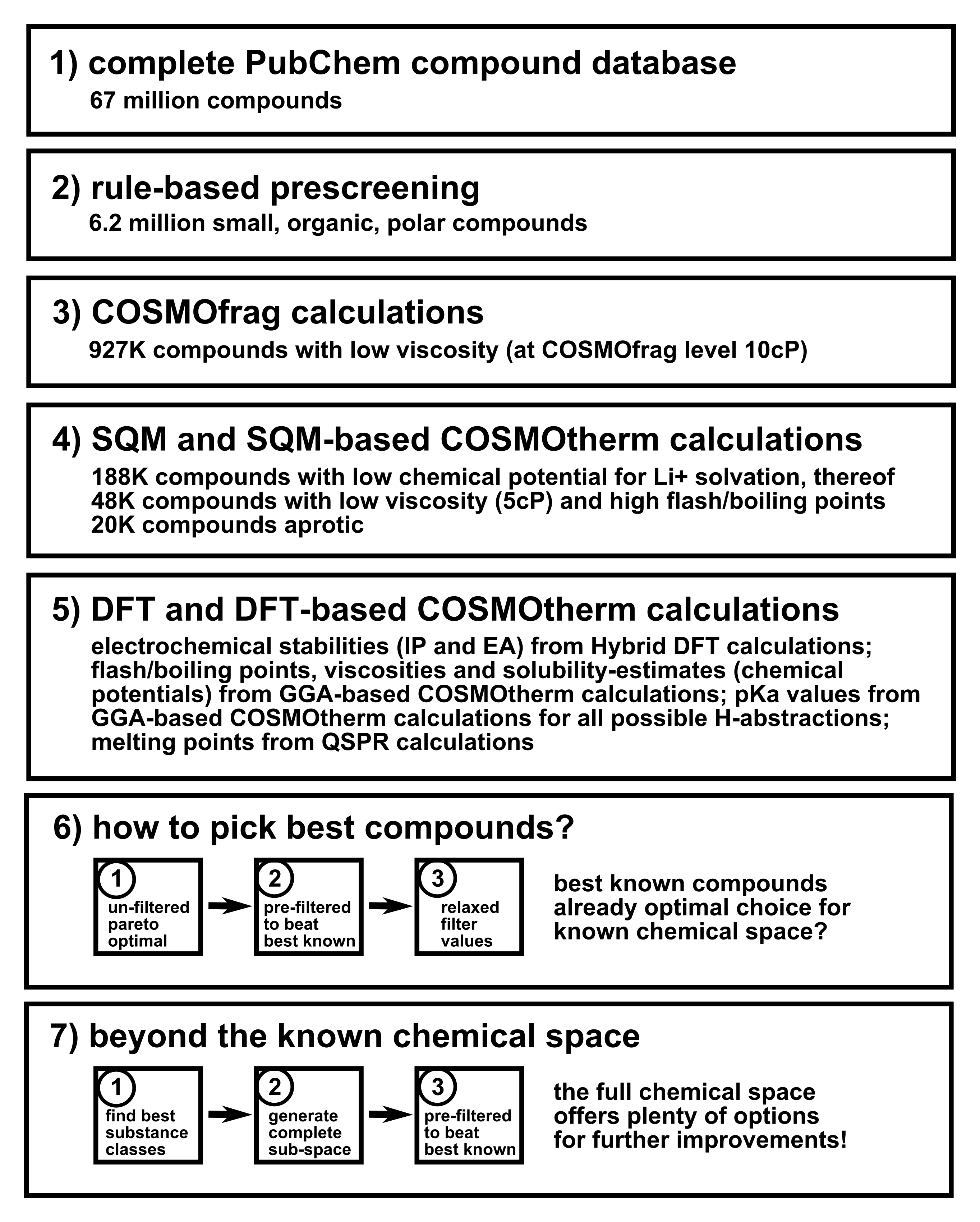}
  \caption{Schematic overview of the screening protocol, see text for details.}
  \label{fig:screeningstrategy}
 \end{center}
\end{figure}

After retrieving and converting structures from the database (step 1),
compounds were pre-screened based on simple rules (step 2):
Candidates with more than 18 heavy atoms or elements other than 1st and 2nd row elements, as well as pure hydrocarbons were excluded.
The remaining 6.2 million structures were subjected to very fast COSMOfrag calculations\cite{COSMOfrag} to evaluate their viscosity (step 3).
All compounds with a viscosity greater than 10.0 cP (at COSMOfrag level) were discarded.
The threshold for the viscosity was selected generously to not discard any compounds prematurely.
The remaining 927 thousand polar, organic liquids were evaluated with respect to the chemical potential for $Li^+$ within the bulk compound
based on fast semiempirical quantum mechanical (SQM) PM6-DH+\cite{Korth2010} and SQM-based COSMOtherm calculations,
leaving 188 thousand electrolyte solvent candidates after discarding all candidates with a chemical potential larger than 10 kcal/mol at this level of theory.
These compounds were screened with respect to their ionization potential (IP), electron affinity (EA), viscosity, boiling point,
flash point and the free energies of solvation and chemical potentials of $Li^+$, $O_2^-$ and $O_2$ in the bulk compound at SQM/COSMOtherm level.
48 thousand compounds with a viscosity below 5.0 cP, boiling points above 373K and flash points above 343K were optimized at GGA-DFT level.
(High boiling points are of importance also because the Li-air battery is open to ambient pressure,
so that long-term stability can be problematic in the case of high vapor pressures.)
All before mentioned collective properties have been re-evaluated with COSMOtherm on the basis of the BP86/TZVP calculations.
Additionally, free energies of solvation and chemical potentials of water and carbon dioxide in the bulk compound were included,
as the solubility of water and carbon dioxide should be very low to promote the use of air or less refined oxygen
and minimize parasitic reactions.\cite{Aetukuri2015} 
IP and EA values have been computed at higher levels of theory (Hybrid-DFT, CEPA\cite{CEPA}) to ensure sufficient accuracy.
The final set of molecules comprises about 20 thousand potential polar, non-protic organic liquids that may be good candidates for application in Li-air batteries.
Additional calculations of pK$_a$ values in DMSO with DFT/COSMOtherm were performed for this data set,
to estimate the chemical stability against nucleophilic attacks, H-abstraction reactions and autooxidation. 
We found DFT/COSMOtherm pK$_a$ predictions to be highly correlated (R=0.99) with results from the best methods available,
but computationally much cheaper ({\it c.f.} Supplementary Information Section 3).
For each compound all possible proton abstractions were considered, the lowest pK$_a$ was picked as the descriptor.
A substantial change of molecular geometry after proton abstraction (e.g. ring opening) was taken as a sign
of unsatisfactory electrochemical and chemical stability.
Finally, QSPR melting point predictions were checked for selected compounds.

Our choice of screening parameters does not include estimators for every thinkable property,
but every parameter previously identified as substantially important is included.
Detailed follow-up investigations can be carried out subsequently for the most promising compounds.
For this purpose, and to allow other researchers to try out different strategies for picking best compounds,
the whole data set will be made available on our project web page.\cite{QAH}

%\textcolor{red}{ %ALERT
Our following analysis will very much concentrate on taking into account the above-mentioned experimental results
on the importance of the role of the different ion solubilities.
Other experimentally working groups might want to question the significance of these results
for the further development of Li-air batteries, but we will show below that also competing
experimentally-derived hypotheses can easily be incorporated as selection criteria within our screening approach.
%}
\\

\noindent {\bf \fontfamily{phv}\selectfont{Computational details}}\\
\noindent Ionization potentials (IPs) were calculated at PM6-DH+\cite{Korth2010}, BP86\cite{BP1,BP2}-D3\cite{D3}/TZVP\cite{TZVP},
B3LYP\cite{B3LYP1,B3LYP2}-D3/TZVP and LPNO-CEPA\cite{CEPA}/aug-def2-TZVPP level,
using MOPAC2012\cite{mopac}, Turbomole 6.4.\cite{Turbomole} and ORCA 3.0.3\cite{Orca},
electron affinities (EAs) were extrapolated from IPs and orbital eigenvalues according to Tozer.\cite{Tozer}
Melting points of selected compounds are estimated with a QSPR model of A. Lang.\cite{Lang}
Viscosities, boiling and flash points, pK$_a$ values in DMSO, free energies of solvation and chemical potentials of various ions and molecules
in bulk candidate compounds were computed with COSMOfrag\cite{COSMOfrag} and COSMOtherm\cite{Klamt2011} using SQM and GGA-DFT level input.
%\textcolor{red}{ %ALERT
For flash point calculations we use a constant COSMO area of 39.23 $A^2$ to enforce a better agreement of absolute values with experimental reference data.
%}
%... see above ...
%For the pK$_a$ calculations, every hydrogen atom in a molecule was abstracted as a proton and also the negatively charged molecule was fully optimized.
%The lowest pK$_a$ was then taken as a estimate of the chemical stability. An assessment of the accuracy of this approach is given in Supplementary Information Section 3. 
%For further methodological considerations see Supplementary Information Section 2.
%HERE
The overall computational effort for this study was about 2 million CPU hours.
Here we mostly relied on standard compute cluster resources,
but work on integrating our Volunteer Computing resources more closely is in progress.
\\

\noindent {\bf \fontfamily{phv}\selectfont{Screening Results}}\\
\noindent We then tested several strategies to analyze our screening results and arrive at good suggestions for subsequent experiments,
putting special emphasis on multi-dimensional evaluation to pay tribute to the underlying multi-dimensional problem.
The most obvious approach is to pick Pareto-optimal candidates out of the final set, which gives 37 candidates ({\it c.f.} Supplementary Information Section 4).
DMSO is among the final candidates, which is a clear success for our screening strategy
as this well-performing compound is successfully picked out of several million others.
The other candidates comprise a large variety of N-hetero-cycles, but when checking melting points they are found to be too high for the majority.
Other structural motifs include imines, amides and ureas, which all share the drawback of high melting points or show low $O_2^-$ solubilities.
The most promising suggestions are different sulfoxides, but the calculated pK$_a$ values are lower than for DMSO,
which additionally has the lowest chemical potential for $O_2^-$.
All candidates share low IP values and good oxygen, water and carbon dioxide solubilities,
leading to the conclusion that water and carbon dioxide have to be excluded from the cells in other ways.

As a second strategy we tested pre-filtering compounds to remove cases with poor viscosities and $Li^+$/$O_2^-$ chemical potentials.
Candidates are sorted out that do not beat MeIm (which itself is beaten by DMSO) with respect to $Li^+$ and $O_2^-$ solubility.
This only leaves 7 structures of which 5 are Pareto-optimal ({\it c.f.} Supplementary Information Section 5).
Among the suggestions is an amine oxide that most likely has a very high melting point.
The next two hits, a dihydro-thiophene oxide and a phosphinic acid ester, are most likely reactive and cannot withstand nucleophilic attacks,
which disappointingly leaves us only with DMSO and MeIm.

This surprising result, that the best known compounds are already optimal (or at least very close to optimal)
choices within the known chemical space (as pre-selected by the filtering) clearly merits further investigation.
A key issue is a low chemical potential for $Li^+$ in combination with a low chemical potential for $O_2^-$,
that is the two-dimensional problem very recently identified by Luntz and co-workers.
Our screening results now show that (even in first approximation) a third dimension needs to be considered:
Low chemical potentials for both ions, i.e. high donor and acceptor numbers
are connected to strong intermolecular interactions in the pure bulk compound, i.e. a high viscosity and a high melting point.
The problem is illustrated in figure \ref{fig:problem}.

\begin{figure}[h]
 \begin{center}
  \includegraphics[width=0.9\textwidth]{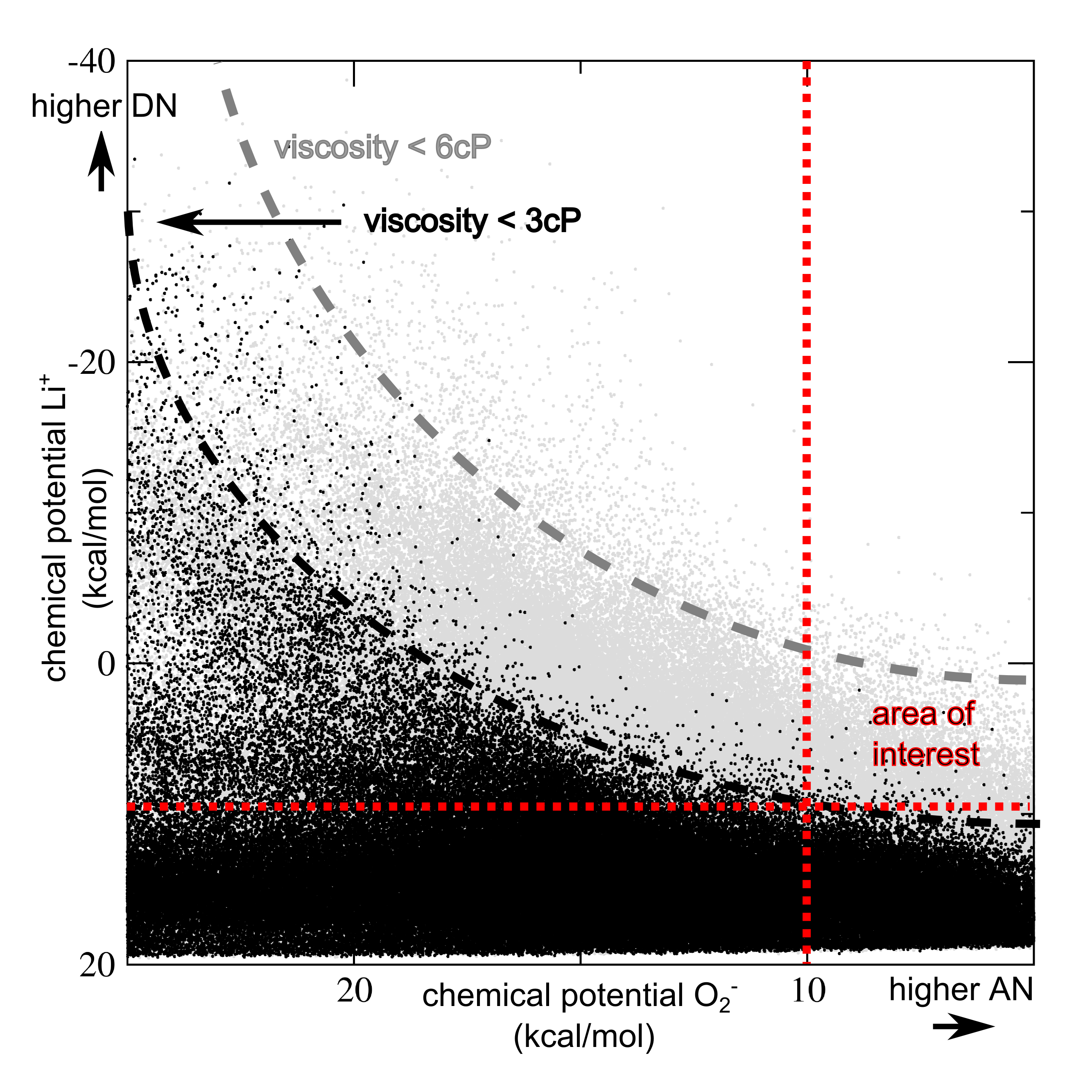}
  \caption{{\bf Illustration of the problem with optimizing both donor and acceptor numbers (DN/AN):}
Chemical potentials for $Li^+$ vs. chemical potentials for $O_2^-$ in the bulk candidate compound
are plotted for all 927.000 compounds treated at SQM level.
($\mu(Li^+)$ is systematically underestimated in comparison to higher-level DFT-based data,
but trends are similar, see Supporting section 1 for details.)
Black dots indicate compounds with a viscosity below 3 cP, grey dots compounds with a viscosity below 6 cP,
the upper right part is empty because of the (pre-)screening step 3.
Black and Grey lines are added to guide the eye.
Lower $\mu(Li^+)$ and $\mu(O_2^-)$ values indicate higher DN and AN,
the search is for compounds with both chemical potentials lower than (approximately) 10 kcal/mol,
indicated by the red, dotted lines.
Almost all compounds in the 'area of interest' have high viscosities (i.e. above 3 cP),
due to strong intermolecular interactions in the bulk solvent.}
  \label{fig:problem}
 \end{center}
\end{figure}

Candidates that clearly beat DMSO and MeIm with respect to both $Li^+$ and $O_2^-$ chemical potentials are highly viscous or solid at room temperature.
The search for new Li-air battery electrolyte solvents should therefore not focus on maximum DN and AN numbers,
but on finding the optimum balance between the two ion solubilities and viscosity.
If a lower performance for one property can be tolerated, room is given to optimize the two other ones.

Given this directive, also compounds with somewhat higher $Li^+$ or $O_2^-$ chemical potentials become interesting.
As a third strategy we therefore used relaxed filter thresholds, adding 10\% to the respective values of MeIm.
The results can be seen in the Supplementary Information Section 6. The biggest share are again N-hetero-cycles with a too high melting point.
Aromatic hetero-cycles with more than one nitrogen in the aromatic rings are especially interesting.
Also sulfoxides are again among the hits, but they again show higher reactivity than DMSO going by the pK$_a$.
The suggestions also comprise phosphine oxides and phosphinic acid esters.

Our study shows impressively how good of a choice DMSO and MeIm are and how strong chemical intuition actually is.
The PubChem database covers, on the other hand, only a tiny fraction of the relevant chemical space
(which we estimate to be at least $10^{10}$ times larger), though due to it's relative homogenity
far less diversity is available than this number suggests.\\

\noindent {\bf \fontfamily{phv}\selectfont{Beyond the known chemical space}}\\
\noindent In the last part of our study we therefore turned to screening full sub-spaces for the most promising compound classes.
These classes were identified by analyzing the average performance of compounds with the same functional groups.
We use Checkmol\cite{Checkmol} to analyze molecules for the presence of various functional groups,
of which 200 different ones are currently implemented in the program.
The most important results of this compound class analysis can be seen in table \ref{tab:fganalysis}
(and with more detail in Supplementary Information Section 7).

\begin{table}[h]
\begin{center}
\caption{Analysis of the performance of the most promising compound classes.$^{1}$}
\label{tab:fganalysis}
\begin{tabular}{p{5cm}cccc}
\hline
Compound Class$^{2}$ & Pic & $\mu(Li^+)$ [kcal/mol] & $\mu(O_2^-)$ [kcal/mol] & Comment\\
\hline
Phosphine Oxide& \includegraphics[trim=0 0 0 -5, scale=0.5]{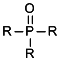} & 5.85 & 13.43 & selected \\
Phosphine&\includegraphics[scale=0.5]{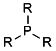} & 6.34 & 13.55 & assumed to be reactive \\
Carboxylic Acid Amidine&\includegraphics[scale=0.15]{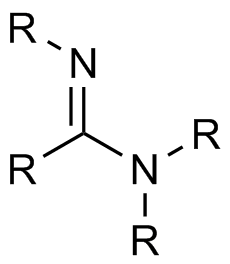}& 8.13 & 15.44 & \\
Secondary Aliphaticaromatic Amine Alkylarylamine& N-Rings & 8.91& 12.65 & selected \\
Guanidine&\includegraphics[scale=0.5]{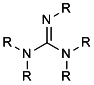}& 9.16 & 16.32 & \\
Sulfoxide&\includegraphics[scale=0.5]{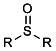} & 9.41 & 12.82 & low pKas \\
Imine&\includegraphics[scale=0.5]{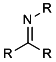} & 10.27 & 15.39 & low pKas \\
Carboxylic Acid Hydrazide&\includegraphics[scale=0.5]{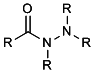}& 10.33 & 14.75 & low pKas\\
Lactam&\includegraphics[scale=0.15]{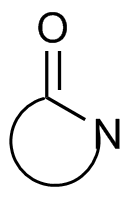}& 10.42 & 14.80 & \\ 
Secondary Amine&\includegraphics[scale=0.5]{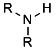}& 10.46 & 15.29 & selected \\
Urea&\includegraphics[scale=0.5]{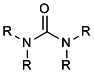}& 10.46 & 15.42 & \\
Oxohetarene&\includegraphics[scale=0.5]{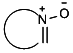}& 10.48 & 12.96 & high viscosities \\
Secondary Aliphatic Amine Dialkylamine& N-Rings & 10.61 & 15.53 & \\
Tertiary Aliphaticaromatic Amine Alkylarylamine & N-Rings& 10.65 & 13.76 & high viscosities\\
Azide&\includegraphics[scale=0.5]{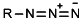}& 10.65 & 14.85 & low pKas\\
Tertiary Carboxylic Acid Amide&\includegraphics[scale=0.5]{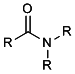}& 10.84 & 14.98 & \\
\hline
\end{tabular}
$^1$ Values averaged over all entries for each compound class, listed are classes with $\mu(Li^+)>$MeIm+10\%,
for more details see Supplementary Information Section 7.\\
$^2$ According to Checkmol classification.
\end{center}
\end{table}

Phosphine oxides show a very good overall performance, with excellent $Li^+$ solubilities, but are known to have high melting points.
This compound class was selected for further investigations to look for a candidate with a good balance between the $Li^+$/$O_2^-$ solubility and viscosity.
Amidines, guanidines and imines are interesting suggestions, but though they show a good $Li^+$ solubility,
the balance with $O_2^-$ is not very promising and the non-aromatic C=N bond is likely reactive.
The second compound class we chose for further investigations were hetero-cycles, especially those containing Nitrogen.
Various functional groups according to Haider fall in this category, for example numerous amines, lactames, and oxohetarenes.
Their properties are over all very promising aside from somewhat high viscosities. 
Sulfoxides show as well a very good balance for the chemical potentials, but the average pK$_a$ is worrying.
As within the PubChem no sulfoxide better than DMSO could be found and a large amount of sulfoxides was already included in the PubChem database,
this compound class was not chosen for further studies here.
(The properties of DMSO may nevertheless be succeeded by sulfoxides with additional functional groups of different type.)
Derivates from carboxylic acids and urea may also be interesting for further investigation, but are not included in this study.

A comparison with the literature supports the validity of our screening results:
%HERE
%Only one compound class that was previously excluded due to experimental or theoretical findings is found in our list of best performing ones.
%Such compound classes would for example be organic carbonates, sulfonates, pure esters, lactones and ethers.
Several compound classes have been identified to fail as Li-air battery electrolyte solvents, for example organic carbonates, sulfonates, pure esters, lactones and ethers.
Only one compound class that was previously excluded due to experimental or theoretical findings is found in our list of best performing ones.
The compounds incorrectly listed are phosphinc acid esters, which were shown to be susceptible to nucleophilic attacks by Bryantsev {\it et al. }.\cite{Bryantsev2013a} Though we do not cover this type of reactivity, pK$_a$s lower than for DMSO and MeIm and thus lower chemical stabilities are predicted.
Our study is further supported by the fact that many compound classes previously identified as promising are among our hits.
Examples are lactams, amides, phosphine oxides and N-hetero-cycles.\cite{Bryantsev2011, Bryantsev2013a, Balaish2014}
We do not list nitriles, as we find comparably poor $Li^+$ solubilities on average, but otherwise reasonably good properties.
%HERE
Our findings also indicate, that when screening existing databases, further reactivity estimates beyond the pK$_a$ need to be included.
Additional estimates may for example take the reactivity of double or ester bonds or highly strained ring structures into account.
These findings can on the other hand easily be incorporated as rules for structure generation when generating new databases,
as we will show in the following.

Based on our analysis of the average performance of compound classes, phosphine-oxides and hetero-cycles mainly containing nitrogen,
but also oxygen and sulfur were chosen for further investigations.
Turning away from the known chemical space of the PubChem database,
we screened the full chemical sub-spaces of these compound classes within certain structural constraints.
The Molgen algorithm\cite{Molgen} was used to construct all possible structures for the relevant sub-spaces.
To keep the number of structures manageable, structure generation was broken down into parts.
For phosphine-oxides for instance we first looked at aliphatic structures and had to keep the overall number of atoms low,
while in the second step we looked at cyclic phosphine-oxides, where a much larger overall number of atoms was possible,
because many atoms were bound to end up in the enforced ring motif.
Turning to hetero-cycles we first looked at mono- and bi-cycles, but constrained to aromatic systems and only considering nitrogen heteroatoms.
As bi-cycles did not give promising results, we turned to 5-6 membered mono-cycles, still only considering nitrogen but now also non-aromatic systems.
To investigate also oxygen and sulfur systems and N/O/S mixed ones we had to turn to constructing simple N/O/S 5- and 6-ring hetero-cycles first
and add aliphatic rests to these core rings later on.

Overall, five different investigations were carried out:
First, all phosphine-oxides $P_1O_1C_{3-6}$ and their sulfur analogs $P_1S_1C_{3-6}$ were constructed with no rings other than 5- to 7-membered ones allowed.
We then applied the screening protocol outlined above for the PubChem database, starting at DFT level (step 5).
Out of 362 compounds we identified 10 Pareto-optimal structure with a promising balance of the chemical potential of $Li^+$ and $O_2^-$ ({\it c.f.} Supplementary Information Section 8). 
Unfortunately all structures with a low melting point are most likely reactive, because they incorporate double/triple bonds or allene structures. 
The most promising structures, an aliphatic phosphinan-oxide and an aromatic phosphol-oxide (ID po138, po922), have an outstanding balance of the chemical potential of $Li^+$ and $O_2^-$ and good safety features, 
but are barely liquid.
In a second run, all cyclic phosphine-oxides $P_1O_1C_{5-10}$ were constructed with
one ring enforced, only 5-6 membered rings allowed, and double or triple bonds except aromatic ones forbidden.
Screening was again started at DFT level with 926 compounds, but all were showing high $\mu(O_2^-)$ values. 
As third step, all aromatic N-hetero-cycles $N_{1-3}C_{2-12}$ were constructed with the same constraints as for the cyclic phosphine-oxides,
i.e. all aromatic N-based mono- and bi-cycles. It should be noted that the structure generator does only count ring-wise fully conjugated
double bonds as aromatic, so that structures like MeIm are not included in this set.
Screening started at DFT level (step 5) with 28356 structures, but no compound turned out to be competitive.
As a forth step all (including non-aromatic) N-hetero-cycles $N_{i=1-3}C_{3-(10-i)}$ were constructed with one ring enforced, only 5-6 membered rings allowed,
but now also double bonds other than aromatic ones allowed.
113140 structures were evaluated at SQM level (step 4), 1290 at DFT level (step 5).
102 candidates are competitive to MeIm, of which 39 are Pareto-optimal,
and 18 of the latter are 'simple' in the sense that they contain no reactive binding motifs (like double or triple bonds outside the ring).
As a fifth step, all simple, unsubstituted 5- and 6-ring hetero-cycles containing up to 3 nitrogen, oxygen or sulfur atoms were constructed (865 structures)
and then all possibilities of attaching up to three carbon atoms to these core rings were evaluated (204695 structures).
458 candidates are competitive to MeIm, of which 74 are Pareto-optimal,
and 13 of the latter are 'simple' in the sense that they contain no reactive binding motifs and only one hetero-atom species.
(Compounds with only one hetero atom species should be more easily accessible to experiment.)
We list all 31 'simple' heterocycle hits in Supplementary Information Section 9,
but estimated melting points indicate that most of these compounds are again not very likely to be liquid at room temperature.

Further experiments are clearly necessary to find out if higher $\mu(O_2^-)$ can be tolerated to allow for lower $\mu(Li^+)$ values at low viscosities.
To make suggestions for the systematic experimental investigation into the optimal balance of $Li^+$/$O_2^-$ solubility and viscosity,
we again turn to the PubChem database, thereby making sure that suggested compounds are (more or less) readily available.
Supplementary Information Section 10 gives a compilation of low-viscosity compounds with very different balances of the two relevant chemical potentials.
From all compounds of the final set with a viscosity lower than 2 cP and a chemical potential for $Li^+$ lower than 10 kcal/mol,
compounds with the lowest chemical potential for $O_2^-$ are given for each 1 kcal/mol interval of $\mu(Li^+)$.

The PubChem and hetero-cycle data is readily available for re-evaluation with adjusted filter thresholds if higher $\mu(O_2^-)$ can indeed be tolerated.
Raw data for all screening runs will accordingly be made available on our project web page to encourage further investigations also by other researchers.
%\textcolor{red}{ %ALERT
As a first example we give a list of the most promising motifs for the case that the solubility of the negative species is actually of lesser importance
(i.e. looking for compounds with a viscosity lower than 2 cP in combination with a Lithium cation chemical potential below 10 kcal/mol,
as well as pK$_a$s higher than 25 and melting points lower than 10 $\,^{\circ}\mathrm{C}$ ) in Supplementary Information Section 11.
Very different compounds are found in this analysis in comparison to the previous ones, now with an emphasis on amide and amine motifs,
thus nicely illustrating how important the choice of selection criteria (and therefore input from experiment) is.
%}
\\

\noindent {\bf \fontfamily{phv}\selectfont{Conclusions}}\\
\noindent Our systematic investigation of the known chemical space represented by the PubChem database indicates that
the problem of finding better Li-air electrolyte solvents is not only about maximizing donor and acceptor numbers,
but about finding the optimal balance of the relevant ion solubilities with the viscosity of the solvent.
The PubChem results and the subsequent exploration of full sub-spaces for the most promising compound classes delivered
a list of compounds for the experimental investigation of this balance.
Our results imply that further trial-and-error investigations of commercially available chemicals are most likely doomed to failure.
Instead the exploration of unknown substances should be pursued, using both computational and experimental screening techniques.
We did for instance not investigate compounds with multiple functional groups apart from those in the PubChem database,
as well as the opportunities offered by mixtures of known and/or unknown solvents,
but both tasks can be well-handled on the computational side with the screening approach proposed here.
We have thus good hopes that supplementing experimental battery research with a theory-based,
rational decision model will help to speed up the transfer of insight into innovation in this field.
\\

\noindent {\bf \fontfamily{phv}\selectfont{Acknowledgments}}\\
\noindent The authors would like to thank Sylvain Brimaud, Gerhard Maas and Andreas Klamt for helpful discussions and
gratefully acknowledge financial support from the Barbara Mez-Starck Foundation.
%HERE
%\textcolor{red}{ %ALERT
This work was supported in part by the bwHPC initiative and the bwHPC-C5 project provided through associated compute services of the JUSTUS HPC facility at the University of Ulm.
bwHPC and bwHPC-C5 are funded by the Ministry of Science, Research and the Arts Baden-W\"urttemberg and the Germany Research Foundation.
%}
\\

%\noindent {\bf \fontfamily{phv}\selectfont{Author ontributions}}\\ \\
%\noindent Both authors contributed equally to all parts of the work.
%\\

\noindent {\bf \fontfamily{phv}\selectfont{Supplementary Information}}\\
\noindent Supplementary Information is available online at ...
Raw data will be made available as a web-accessible database on http://qmcathome.org/clean\_mobility\_now.html.
\\

\bibliographystyle{achemsolx2}
\fontfamily{phv}\selectfont{\bibliography{bib_paper}}

}%Helvetica
\end{document}